\documentclass[conference]{IEEEtran}
\usepackage[T1]{fontenc}
\usepackage{amsmath,amssymb}
\usepackage{booktabs}
\usepackage{graphicx}
\usepackage{xcolor}
\usepackage[hidelinks]{hyperref}
\usepackage{listings}
\newcommand{\ftinj}{\textsc{ftinj}}
\newcommand{\gamess}{\textsc{gamess}}

\begin{document}

\title{Validating LLM-Modernized Scientific Software Through Differential Fault Injection}

\author{%
\IEEEauthorblockN{Evan Coleman}
\IEEEauthorblockA{University of Mary Washington\\
Fredericksburg, VA, USA\\
ecolema4@umw.edu}
\and
\IEEEauthorblockN{Yuzhong Shen}
\IEEEauthorblockA{Old Dominion University\\
Norfolk, VA, USA\\
yshen@odu.edu}
\and
\IEEEauthorblockN{Masha Sosonkina}
\IEEEauthorblockA{Old Dominion University\\
Norfolk, VA, USA\\
msosonki@odu.edu}
\and
\IEEEauthorblockN{Peng Xu}
\IEEEauthorblockA{
Ames National Laboratory \\
and Iowa State University}
Ames, IA, USA \\
pxu@iastate.edu
}

\maketitle

\begin{abstract}
Large language model (LLM) coding agents are increasingly being used
to modernize the legacy Fortran that underlies production scientific
software.  Validation of these transformations generally emphasizes
nominal executions, however, and therefore may not determine whether
a modernized implementation preserves the response of the original code 
to faults, numerical perturbations, and reduced precision.  We
present differential fault injection as a validation method for
LLM-modernized scientific software.  The proposed harness instruments
the shared self-consistent-field (SCF) driver of \gamess{} at twelve
sites and applies identical, deterministic faults to the original and
modernized implementations.  This placement isolates the converted
integral kernels while avoiding differences in injection location or
timing.

Across more than 2{,}200 runs, the campaigns characterize transient,
persistent, parallel, and precision-related perturbations.  For
transient faults, the measured absorption costs are consistent with a
contraction-based model evaluated at two contraction rates: predicted
slopes of 0.74 and 1.49 additional iterations per bit correspond to
measured slopes of 0.82 and 1.50.  Persistent Fock perturbations
produce an approximately factor-of-two reduction in final-energy error
per additional bit and place the tested silent-data-corruption
transition near $2^{-22}$.  Additional campaigns identify
phase-dependent parallel deadlocks, convergence to alternate SCF
solutions following large orbital perturbations, and false convergence
under reduced Fock precision.  The parallel experiments also evaluate
a one-line synchronization change suggested by an existing developer
comment; within the tested campaign, the change eliminates the observed
deadlock channel while preserving outcomes outside its intended scope.

For the primary differential experiment, the original and modernized
kernels produce identical outcome classifications, iteration counts,
injection counts, and final energies for 200 paired injections.  After
composing the synchronization change with the modernization and
matching the harness and compilation configurations, the resulting
implementation also matches the hardened original in all 40 paired
runs.  These results support differential fault injection as a
practical method for evaluating off-nominal fidelity and targeted
resilience improvements in AI-assisted scientific-software
modernization.
\end{abstract}
 
 
\section{Introduction}
\label{sec:intro}
 
Many production scientific-computing packages rely on legacy code
bases.  Packages such as
\gamess{}~\cite{schmidt1993general,barca2020recent,zahariev2023general}
contain decades of Fortran development in which numerical algorithms,
performance optimizations, and resilience mechanisms are closely
integrated.  Maintaining and modernizing these implementations requires
specialized knowledge of both the application domain and the existing
software architecture.  LLM coding agents provide one possible approach
to this task through supervised, incremental translation of selected
kernels into modern Fortran under explicit style and testing
requirements.  The modernization effort considered here follows this
approach: an LLM agent converts low-level integral-evaluation files to
free-form Fortran within a curated workflow, with the standard
\gamess{} validation suite applied after each change.

This study evaluates behavior outside nominal executions.  A passing
test suite establishes correctness for the inputs and execution
conditions represented by that suite, but generally does not exercise
the response of the implementation to numerical perturbations or
partial corruption of its internal state.  Iterative solvers may
attenuate perturbations, while convergence safeguards,
symmetrization, damping, and re-orthogonalization may reduce their
effects or convert them into additional iterations.  Other
error-handling paths are executed only after an abnormal condition
has occurred.  An LLM modernization may preserve nominal outputs
without necessarily preserving these off-nominal behaviors. This off-nominal
behavior is precisely the kind of property that current AI-code
acceptance practice does not measure, and, as faults in large
systems become more frequent and more
subtle~\cite{dixit2021silent,hochschild2021cores}, it is the property
on which trust in the modernized code ultimately rests.
 
We evaluate this question using \emph{differential fault injection}.
Identical, reproducible faults are introduced into the original and
LLM-modernized implementations, and the resulting executions are
compared on a paired basis.  The injection sites are placed in driver
code shared by both implementations.  In the present study, the sites
are located in the \gamess{} self-consistent-field (SCF) iteration at
the interfaces to the converted kernels.  A perturbed density matrix
therefore enters either the original integral implementation or its
LLM-modernized counterpart through the same driver-level site.  This
design reduces differences in injection placement and timing and
isolates the converted kernels as the varying component.

Throughout this paper, we distinguish \emph{off-nominal fidelity}
from \emph{absolute resilience}.  The differential experiments ask
whether the modernization preserves the response of the original implementation
observed within a specified fault campaign.  The original
implementation therefore serves as the reference for fidelity, but
its behavior is not assumed to be desirable.  Separate hardening
experiments evaluate whether a localized weakness can be reduced
without introducing regressions elsewhere in the tested fault space.
 
The same instrument also supports an \emph{agentic hardening loop}.
Campaign results are used to localize specific weaknesses, such as a
convergence scalar for which replicas do not establish agreement.
Each finding is then provided to the LLM agent as a bounded repair
task within the same curated workflow used for modernization.  The
resulting change is accepted only after the campaign is repeated
against explicit criteria: no new silent data corruptions relative
to the reference implementation, fewer hangs, and no degradation in
the absorption behavior of runs that complete under both versions.
The human developers define the fault model and acceptance criteria;
the agent proposes the implementation change; and the experimental
campaign evaluates the resulting behavior.  We use this
measure--localize--modify--re-measure process as a case study in
measurement-guided AI-assisted maintenance of scientific software.
 
The principal contributions of this paper are:
\begin{itemize}
\item \textbf{A differential validation protocol for
LLM-modernized scientific kernels.}
Identical faults are introduced through driver code shared by the
original and modernized implementations, allowing paired comparison
of their off-nominal behavior while isolating the converted kernels
as the unit under test (Sec.~\ref{sec:method}).

\item \textbf{\ftinj{}, an application-aware fault-injection harness
for the \gamess{} SCF iteration.}
The harness provides twelve injection sites, magnitude-parameterized
faults, phase-aware triggers, deterministic per-rank replay, and
run-level provenance while requiring only a small source modification
to the shared SCF driver (Sec.~\ref{sec:harness}).

\item \textbf{A theoretical and experimental characterization of
SCF fault absorption.}
A contraction-based model predicts the additional iterations required
to absorb transient perturbations.  Experiments on two systems with
different measured contraction rates produce slopes consistent with
the corresponding predictions.  Additional campaigns characterize
persistent faults, orbital perturbations, reduced precision, and
replica-local faults (Secs.~\ref{sec:model} and~\ref{sec:results}).

\item \textbf{A paired evaluation of an LLM-modernized integral
kernel.}
Across 200 paired injections, the original and modernized
implementations agree in outcome classification, iteration count,
injection count, and final energy.

\item \textbf{A measurement-guided hardening case study.}
The parallel campaign additionally motivates and evaluates a
localized synchronization change to the SCF convergence logic.
Within the tested campaign, the change eliminates the observed
deadlock channel and composes with the modernized kernel without
introducing paired outcome differences.
(Sec.~\ref{sec:endstate}).
\end{itemize}
 
\section{Background and Related Work}
\label{sec:background}
 
\subsection{\gamess{} and Self-Consistent Field Iteration}
\gamess{}~\cite{schmidt1993general,barca2020recent} is a widely used
general-purpose quantum chemistry package whose core is written in
fixed-form Fortran.  The restricted Hartree--Fock (RHF) path solves
the self-consistent field equations by fixed-point iteration: from a
density matrix $D_k$, build the Fock matrix $F(D_k)$ (the dominant
cost, via two-electron integral evaluation); optionally accelerate
(DIIS interpolation~\cite{pulay1980convergence,pulay1982improved},
second-order orbital rotations, damping, extrapolation, level
shifting); diagonalize to obtain orbitals; form $D_{k+1}$; and test
convergence on the density change and energy change.  Two properties
of this structure matter for resilience.  First, the iteration is
contractive near the solution, so transient perturbations are
attenuated geometrically, which is the classic argument that iterative
methods carry intrinsic fault tolerance.  Second, the production
converger is a \emph{stateful mechanism} (one-way switches between
accelerators, history-dependent interpolation, consecutive-pass
convergence tests), so the response to a fault depends on \emph{when}
in the convergence trajectory it lands, not only on its size.
 
The Hartree--Fock energy surface introduces an additional
consideration: the
Hartree--Fock energy surface over orbital choices has multiple
stationary points, and ``converged'' certifies self-consistency,
not identity of the solution reached.
 
\subsection{Soft errors in computational chemistry}
van~Dam et al.~\cite{van2013case} injected bit flips into
Hartree--Fock in NWChem and found that while many corruptions are
absorbed, algorithmic self-correction alone is not a sufficient
defense; related NWChem work explored fault-tolerant execution via
over-decomposition, selective replication of read-only versus
read-write data, and resilient Global
Arrays~\cite{daily2014suitability}.  We believe that  no comparable
classical fault-injection characterization exists for \gamess{}, prior 
work has not, to our knowledge, examined whether a semantically equivalent 
source transformation preserves the
application response to injected faults.
 
\subsection{Fault-tolerant iterative methods}
The numerical-methods community has developed both theory and
mechanism for iteration-level resilience: selective reliability, in
which only a small critical fraction of the computation is executed
reliably~\cite{hoemmen2011fault}; skeptical programming and
bounded-error checks exploiting mathematical invariants of the
solver~\cite{elliott2014evaluating}; and fault-tolerant variants of
asynchronous fixed-point iterations, including analysis of when
convergence survives lost or stale
updates~\cite{coleman2021fault,coleman2026fault}.  The DIIS accelerator at the heart
of SCF is a specialization of Anderson
acceleration~\cite{walker2011anderson}, which connects the present
study to that body of fixed-point fault-tolerance results and
underlies the contraction-cost model of Sec.~\ref{sec:model}.
 
\subsection{Fault Models and Validation of AI-Transformed Software}
Hyperscaler fleet studies established that silent data corruption
from marginal or ``mercurial'' cores is a real, recurring
phenomenon at scale~\cite{dixit2021silent,hochschild2021cores},
sharpening a concern the HPC resilience community had projected for
exascale systems~\cite{snir2014addressing}.  These observations
motivate our fault-model emphasis on magnitude-parameterized and
sticky corruptions rather than single-bit transients alone, and its
unification of fault magnitude with limited-precision arithmetic.  
LLMs are also increasingly used to generate and transform
code~\cite{liu2023your,zheng2023survey,nitin2024using,sanas2025llm}.
Validation of these transformations remains primarily based on test
suites, compilation, and human review.  The present work augments
these methods with paired measurements of off-nominal behavior and
evaluates whether the same procedure can be used to assess a
subsequent resilience modification.
 
\section{Differential Validation Methodology}
\label{sec:method}
 
\subsection{Fault Model and Validation Objective}
We consider transient and sticky data corruptions in the floating-point
state of a running SCF calculation (application-level proxies for bit flips, 
marginal-core arithmetic errors, and precision loss) plus
timing perturbations.  We deliberately parameterize faults by
\emph{magnitude in equivalent bits}: a relative perturbation
$\eta = 2^{-B}$ applied to selected elements, so that $B{=}52$
approaches roundoff, $B{=}24$ mimics single-precision-level error,
and smaller values of $B$ correspond to larger perturbations.  Bit flips are retained as
a mode (sign/exponent/mantissa-targeted), as are truncation modes
that emulate fp32/bf16/fp16 storage of a chosen array. This parameterization allows 
hardware-inspired perturbations and
reduced-precision storage experiments to be compared using a common
magnitude scale.
 
The primary validation question is comparative: for each paired
injection in the specified campaign, do the original and
LLM-modernized implementations produce the same outcome
classification, iteration count, and final numerical result?
Aggregate outcome distributions are also reported, but the paired
comparison is the principal acceptance criterion because identical
marginal distributions can conceal disagreements between individual
runs.  The campaign therefore evaluates off-nominal fidelity within
the tested sites, magnitudes, phases, and configurations; it does not
establish equivalence for all possible faults or executions.
Absolute resilience of \gamess{} is characterized separately and is
of independent interest.
 
\subsection{Shared-Driver Instrumentation}
The modernization under study converts the lowest-level two-electron
integral files to modern Fortran while leaving the SCF driver
(\texttt{rhfuhf.src}) untouched.  We exploit this: \emph{every}
injection site is placed in the shared driver, at the data interfaces
surrounding the converted code.  Corrupting the density immediately
before the Fock build (site S1, Table~\ref{tab:sites}) feeds
identical corrupted input to either the legacy integral path or its
LLM translation; all downstream sites observe the consequences.  The
identical fifteen-line patch applies verbatim to both trees; this common 
patch reduces the possibility that differences in site
placement affect the paired comparison. The converted kernels are the
only varying component, which is the  unit under
test of the differential experiment.  The installer locates each site by code
\emph{anchor pattern} scoped to the target subroutine rather than by
line number, verifies that each pattern has a unique match, and terminates with an
error if the expected source context has changed.
 
The unit under test is also isolated at run time.
Conventional SCF computes integrals once, outside the iteration;
the differential decks therefore select direct SCF
(\texttt{DIRSCF}), placing the integral code inside the faulted
loop each iteration, and force every shell quartet through the
general Rys path (\texttt{INTTYP=RYSQUAD}), so the converted
kernels perform all integral work and density corruption perturbs
their control flow (screening decisions) as well as their inputs.
 
\subsection{Outcome Classification and Mechanism Attribution}
Each run is classified against a cached unfaulted baseline into seven
classes: \textsc{absorbed-clean} (converged, $|\Delta E| \le$ tol,
no extra iterations), \textsc{absorbed-delay} ($k$ extra iterations:
the \emph{cost} of absorption), \textsc{sdc} (converged with a 
final-energy difference exceeding the
specified tolerance), \textsc{bad-energy},
\textsc{unconverged}, \textsc{crash}, and \textsc{hang}
(wall-clock timeout with process-group kill).  The experiments 
also record the activation of existing \gamess{}
response mechanisms, including diagonalization-error termination,
the two-consecutive-pass convergence rule, damping following energy
increases, DIIS resets, and Fock symmetrization.  These observations
are used to associate absorbed faults with specific mechanisms where
possible.
 
Attribution is sharpened by running every campaign in two input-deck
configurations requiring zero code changes.  Config~A uses the 
production-default convergence configuration.
Config~B (bare fixed point: DIIS, second-order convergence,
damping, extrapolation, and level shifting all disabled) isolates raw
Roothaan contraction, which is the regime where fixed-point fault-tolerance
theory applies cleanly.  The A$-$B delta assigns absorption credit to
specific accelerator machinery. The comparison is used to estimate the contribution of the
acceleration machinery to the observed response.
 
The campaigns use two element-selection policies: a uniformly random
element (average case; on small high-symmetry systems this
occasionally lands on structural zeros, and such \emph{null
injections} are identified post hoc from the harness log and reported
as their own absorption channel) and the largest-magnitude element (a deterministic
maximum-magnitude policy that avoids null injections).
 
The implementation of Config~B requires one clarification. \gamess{}
does not ordinarily permit a nontrivial molecular calculation to proceed
without an SCF accelerator. If both DIIS and SOSCF are disabled, the
input-processing logic automatically re-enables an accelerator unless
the documented \texttt{NOCONV=.TRUE.} option is specified. This option
is set automatically for systems with fewer than ten basis functions,
so the smaller test system followed the intended unaccelerated path
without further modification. The larger system required the keyword
to be supplied explicitly. We identified this difference by comparing
the converger state reported by \gamess{} with the configuration
specified in the input deck. This behavior is relevant to the present
study because it illustrates an additional form of defensive logic that
must be accounted for when constructing controlled fault-injection
experiments.
 
\subsection{Measurement-Guided Hardening Workflow}
\label{sec:loop}
The differential protocol validates the \emph{translation} done by AI; the
same instrument then supervises the AI-made \emph{repairs}.  Each
campaign finding that localizes a weakness is converted into a
bounded task for the LLM agent inside the modernization workflow
(style guide, review, test gates), for example: ``promote the
master-rank broadcast of the convergence scalars from GDDI-only to
all parallel runs'' or ``add an invalid-value guard on the Fock
matrix before diagonalization.''  The relevant campaign is then repeated.  
A proposed change is accepted
only if it satisfies the following prespecified criteria:
\emph{(i)} it introduces no new \textsc{sdc}, \textsc{crash}, or
\textsc{hang} outcomes relative to the reference implementation;
\emph{(ii)} it reduces the number of outcomes associated with the
targeted failure mechanism; and \emph{(iii)} among runs that complete
under both implementations, it does not increase the measured
absorption cost outside a specified tolerance.  This division of
responsibility is intentional: the human developers define the
validation campaign and acceptance criteria, the agent proposes the
source modification, and the resulting implementation is evaluated
experimentally rather than accepted on the basis of review alone.
Section~\ref{sec:prelim} identifies the first target evaluated through
this process.
 
\section{FTINJ: An Application-Aware Fault-Injection Harness}
\label{sec:harness}
 
\ftinj{} is a self-contained fixed-form Fortran module (no library
dependencies; an internal xorshift generator supplies all
randomness) plus an anchor-based installer and a campaign driver.
The design objectives are behavioral transparency when no site is
enabled, deterministic replay, and a limited source-code footprint.
 
\subsection{Injection Sites}
Table~\ref{tab:sites} lists the twelve sites, chosen to cover the SCF
state space: initial state, the density--integral interface, the Fock
matrix before and after symmetrization and after the converger, the
orbital update (a merge point through which both the diagonalization
and second-order paths pass every iteration), the three
convergence-decision scalars, and the converged results as consumed
downstream.  The post-convergence sites (S6) address a subtlety of
persistence: downstream \gamess{} modules read orbitals and density
from the dictionary file rather than memory, so S6 re-saves the
corrupted state through the checkpoint routine in the code (guarded
so that baseline behavior is untouched) to guarantee the corruption
actually propagates to consumers.
 
\begin{table}[t]
\caption{\ftinj{} injection sites (all in the shared SCF driver).}
\label{tab:sites}
\centering
\scriptsize
\begin{tabular}{@{}lll@{}}
\toprule
Site & Quantity & Placement / role \\
\midrule
S0  & initial density        & before loop entry; guess sensitivity \\
S1  & density $D_k$          & top of iteration; \emph{input to
                               (converted) integral code} \\
S2P & Fock $F$               & after parallel sum, before
                               symmetrization \\
S2  & Fock $F$               & after symmetrization; enters converger \\
S3  & Fock $F$               & after DIIS/damping/shift; enters
                               eigensolver \\
S4  & orbitals $C$           & before density build; both converger
                               paths \\
S5$\times$3 & DIFF, $\Delta E$, DIIS err & convergence scalars;
                               8-byte SDC targets \\
S6$\times$3 & final $C$, $D$, $E$ & post-convergence, persisted to
                               DAF; downstream probe \\
\bottomrule
\end{tabular}
\end{table}
 
\subsection{Fault Modes, Triggers, and Reproducibility}
Ten modes: relative and absolute noise (magnitude $\eta=2^{-B}$),
targeted bit flips (sign / exponent / mantissa / uniform / explicit
bit), whole-array truncation to fp32/bf16/fp16 storage, special
values (NaN, Inf, zero), and a busy-wait delay, which changes execution timing without modifying
the numerical state and is also used to verify propagation of the
environment configuration. Triggers: a specific iteration; sticky from an
iteration onward (the application-level analogue of a mercurial
core); every iteration; or \emph{phase-triggered}, that is to say armed when the
density change first drops below a threshold, so that ``early'' and
``near-converged'' mean the same thing across molecules with
different iteration counts.  Configuration is environmental
(nine variables), injections are capped by a hit budget, and the
random stream is seeded and decorrelated per rank: identical settings
reproduce identical injections across runs, which we verify by
byte-identical event logs across independent invocations.  Every
firing logs the site, iteration, rank, element index, and old/new
values to a per-rank file, giving each campaign row a complete
provenance trail (and making null injections identifiable).
 
\subsection{Instrumentation Footprint and Dormant-Harness Validation}

The source modification consists of fifteen inserted lines at eight
locations in one file and one additional object file in the build.
The patch is installed and removed by an idempotent anchor-based
tool.  When no site is enabled, each hook performs one logical test
and does not modify the application state.

We evaluated the dormant harness using the standard \texttt{exam01}
geometry-optimization case.  The patched original and independently
compiled modernized implementations both reproduced the reference
trajectory: seven SCF cycles, energies identical to all ten printed
decimal places, and per-cycle iteration counts of
$8,7,7,8,5,5,3$.  No harness events were recorded.  We subsequently
ran the 49 modernization-project  cross-path integration
battery on the composed implementation described in
Sec.~\ref{sec:endstate}.  All tests completed normally with
floating-point exception trapping enabled.  After removing
wall-clock-dependent output, 48 tests were byte-identical to their
cached baselines.  The remaining test differed by one unit in the
sixth significant figure of an intermediate response-solver error
norm; all reported final energies, gradients, and geometries were
identical to the available printed precision.  The calculations used
single-threaded reference BLAS and LAPACK to reduce variability in
the experimental baselines.
 
\subsection{Campaign Execution and Provenance}
A dependency-free driver runs a cached unfaulted baseline, sweeps the
cartesian fault space (site $\times$ mode $\times$ magnitude
$\times$ phase $\times$ element policy $\times$ seed) with each run
in an isolated directory and scratch space, kills hung runs by
process-group timeout, classifies per the taxonomy, and emits CSV.
Runs are self-cleaning on directory reuse. 
 
\section{Contraction Model for Transient Fault Absorption}
\label{sec:model}
 
For config~B the SCF is a fixed-point iteration whose contraction
rate $\rho$ can be estimated from the unfaulted density-change
trajectory. 
For the tested configurations, CH$_2$/STO-2G contracts monotonically
with an estimated asymptotic rate $\rho=0.390$, whereas
methanol/6-31G exhibits a decaying oscillatory mode whose tail has an
estimated geometric-mean rate $\rho=0.627$. For the present experiments, 
CH$_2$/STO-2G is treated as a closed-shell singlet RHF test case to exercise 
the RHF SCF path; this choice is not intended to represent the triplet ground state of methylene. 
When modeling a single-shot fault of relative size
$\delta = c\,2^{-B}$ injected when the residual sits at level
$r$, the excess decays at rate $\rho$, a perturbation below the 
ambient residual is not expected to increase
the iteration count.  Hence
\begin{equation}
\Delta k \;\approx\; \max\!\Bigl(0,\;
\frac{\log(\delta / r)}{\log(1/\rho)}\Bigr)
\;=\; \max\bigl(0,\; \sigma\,(B^{*}-B)\bigr),
\label{eq:cost}
\end{equation}
with slope $\sigma = \ln 2/\ln(1/\rho)$ and crossover $B^{*}$
absorbing the order-unity transfer constant $c$.  The slope is
parameter-free once $\rho$ is measured, and the two systems yield distinct 
predicted values: $\sigma = 0.736$ iterations
per bit for CH$_2$ and $\sigma = 1.49$ for methanol (CH$_3$OH).  Both slopes 
were computed from the unfaulted trajectories before
fitting the fault-response data.  The subsequent comparison therefore
tests the predicted dependence on $\rho$ rather than estimating the
slope from the fault campaign itself.
 
Two additional hypotheses were specified before the campaigns.
First, the two-consecutive-pass convergence rule of the code introduces
a discrete contribution to the observed absorption cost: a fault that
interrupts an otherwise successful convergence sequence can add up to
two iterations independently of the contraction rate.

Second, consider a persistent perturbation of magnitude
$O(2^{-B})$ applied at each iteration.  Under a local linearization
of the fixed-point map, the resulting displacement of the fixed point
is also expected to be $O(2^{-B})$.  If the reported energy has a
nonzero first-order response to the perturbed quantity at the selected
injection site, this yields the empirical scaling hypothesis
\begin{equation}
|\Delta E| \propto 2^{-B}.
\label{eq:persistent-scaling}
\end{equation}
The corresponding prediction is a factor-of-two reduction in energy
error for each additional bit of effective precision.  The experiments
in Sec.~\ref{sec:results} test this scaling and identify where the
observed response departs from the local regime.
 
\section{Harness Validation and Motivating Observations}
\label{sec:prelim}
 
Initial validation runs identified three behaviors that informed the
design of the subsequent campaigns.  They also provided checks that
the harness could reproduce {phase-,} {rank-,} and module-dependent
effects.
 
\subsection{Phase-Dependent Fault Absorption}
During the initial geometry-optimization tests, the effect of an
otherwise identical fault depended on the phase of the SCF trajectory.  
An adversarial variant (largest-magnitude
element) reproduced the effect in a single serial run.  
A single-element relative perturbation with $B=24$, introduced at
iteration 3 of each SCF solve, produced no additional iterations in
the early geometry steps but added two iterations in several
near-converged final steps.  In the latter cases, less contraction
remained before termination and the perturbation interrupted the
two-pass convergence sequence.  An adversarial largest-element
injection reproduced the same behavior in a serial run.
 
\subsection{Replica Divergence and Parallel Deadlock}
The \gamess{} SCF driver includes an existing broadcast of the
master-rank convergence scalars for one parallel execution mode.
A nearby source comment attributes previously observed parallel
hangs to small numerical differences between replicas and suggests
extending this synchronization to all parallel runs.  The broadcast
is therefore present in the implementation but is not enabled for
the plain parallel configuration considered here.
The harness allows this proposed failure mechanism to be evaluated
under controlled conditions.  In a four-rank run, a fault applied
to one rank produces a disagreement in the locally computed density
change at the convergence decision.  The clean ranks leave the SCF
iteration, while the perturbed rank enters an additional collective,
causing the execution to deadlock.  This observation motivates
evaluating the broader convergence-scalar synchronization in
Sec.~\ref{sec:hardening}.
The corrupted master-rank iteration log records the injected fault, the 
resulting energy excursion, the subsequent recovery, and the rank-local 
convergence decision. This finding fixes the first repair task of the agentic 
loop (Sec.~\ref{sec:loop}): promote the existing broadcast to all parallel
runs, which is a one-line change whose before/after effect the campaign can
measure directly (\textsc{hang} $\rightarrow$ \textsc{absorbed}/
\textsc{detected}).
 
\subsection{Downstream Propagation Across Module Boundaries}
In the same experiment prefix, before the deadlock, the
single-rank corruption escaped the SCF entirely: the contribution of the perturbed
replica to the globally summed gradient nudged the
geometry optimizer onto a microscopically different trajectory
(energies drifting at the $10^{-9}$ level in subsequent steps).
Faults need not be either absorbed or fatal; they may propagate across
module boundaries at magnitudes below the tolerances used by nominal
regression tests.  This behavior motivates the S6
downstream-propagation sites and illustrates a class of off-nominal
effects that clean-path testing does not exercise.

\section{Results}
\label{sec:results}
 
The campaigns total over 2{,}200 fault-injection runs: 840
single-shot serial runs on CH$_2$ (both configurations), 840 on
methanol, a 400-run differential (both binaries, both
configurations, identical seeds), an 83-run parallel divergence
sweep with follow-ups, a 40-run hardening re-measurement, and a
persistent-fault and precision arm.  All classifications are
against cached unfaulted baselines with an SDC threshold of
$10^{-6}$\,Ha.  Serial single-shot faults were universally
absorbed on CH$_2$; every catastrophic or silent outcome reported
below required persistence, replica asymmetry, gross orbital
corruption, or direct-SCF fault retention. 
No \textsc{sdc}, \textsc{crash}, or \textsc{hang} outcome occurred
for a serial single-shot CH$_2$ perturbation. Such outcomes were
observed only under persistent faults, rank-asymmetric faults, large
orbital perturbations, or direct-SCF fault retention. One
bookkeeping note: the random-element arm on CH$_2$ was degenerate
(symmetry zeros compounded by a harness seed-correlation defect,
fixed in v1.1 and disclosed in Sec.~\ref{sec:threats}); the one-carbon (C$_1$) 
second molecule restored it. The quantitative comparisons 
below therefore emphasize the largest-magnitude-element policy.
 
\begin{figure}[t]
\centering
\includegraphics[width=\columnwidth]{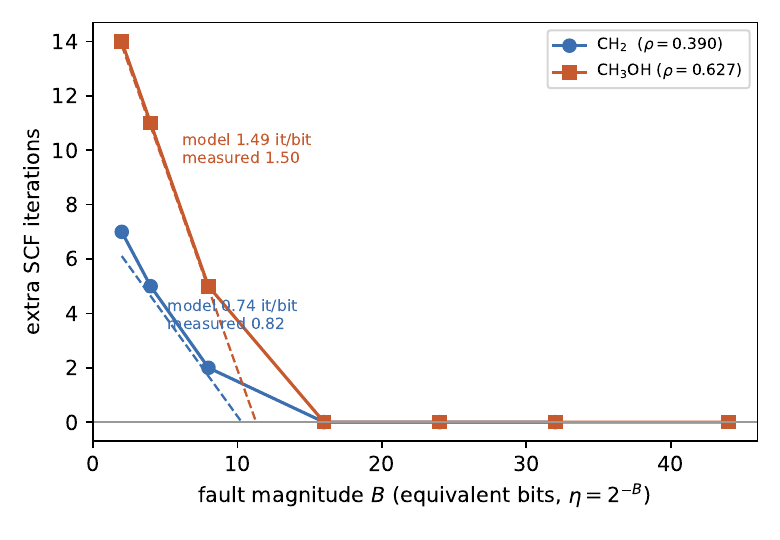}
\caption{Additional iterations as a function of fault magnitude for
near-converged, single-shot Fock-side perturbations under Config~B.
Dashed lines show Eq.~\eqref{eq:cost} using slopes computed from  unfaulted contraction rate 
of each  molecule.}
\label{fig:slope2}
\end{figure}
 
\subsection{Transient-Fault Absorption at Two Contraction Rates}
Figure~\ref{fig:slope2} compares the measured absorption costs with
the slopes specified from the unfaulted contraction rates.  CH$_2$
($\rho=0.390$): registered slope $0.736$, measured $0.82$ over
$B \in \{2,4,8\}$ with zero seed variance, crossover
$B^{*} = 10.3$ recovering the transfer constant at
$c \approx 2.3$.  Methanol ($\rho=0.627$): registered $1.49$,
measured $1.50$ at the density site over three nonzero magnitudes
(the two-point fit of the Fock site gives $1.60$, consistently).  The 
predicted slope is approximately twice as large for methanol as
for CH$_2$, and the measured methanol slope differs from its
prediction by less than one percent.  Boundary behavior matched the
registered corollaries: outcomes quantize in steps of the two-pass
rule, and a handful of gross-magnitude cells \emph{accelerated}
convergence by one iteration (all seeds, deterministically), suggesting that, 
in these cells, the perturbation moved the iterate
onto a trajectory requiring one fewer iteration; these mark 
the linear-response boundary the model
declines to cover.
 
\subsection{Orbital Perturbations and Solution-Space Sensitivity}
The orbital-matrix site S4 produced three different verdicts from
one fault policy, and together they are a single mechanism.  On
CH$_2$, adversarial S4 faults cost exactly zero at every magnitude
and phase in both configurations: the harness provenance logs
show every injection landing at flattened indices 44--45 of the
$7\times 7$ orbital matrix, column 7, the highest \emph{virtual}
orbital ($N_A=4$), and the density build consumes occupied columns
only.  On methanol (C$_1$, $N_A=9$), the largest coefficient sits
in the \emph{occupied} space, and under the production converger
the same faults produced silent corruption in 29 of 30 runs at
$B \le 8$: median final-energy error 13\,mHa, maximum 1.72\,Ha,
values consistent with convergence to alternate SCF stationary
points, despite normal termination.  Under the bare iteration on the same
molecule, the same perturbations were absorbed in 70 of 70 runs:
the bare Roothaan iteration returned to the original solution in all
70 runs, while SOSCF takes
the kicked orbitals as the starting point of its quasi-Newton
rotation and efficiently transports them to a different solution.
These results indicate that the SOSCF update can transport the
perturbed orbitals into a different basin of attraction. The transition 
between alternate and reference final energies occurs
between the tested values $B=8$ and $B=16$.  Vulnerability at the eigenvector site is
therefore entirely a question of which subspace the fault lands in
and which machinery carries it. Neither property is represented directly in the scalar convergence
criterion.
 
\begin{figure}[t]
\centering
\includegraphics[width=\columnwidth]{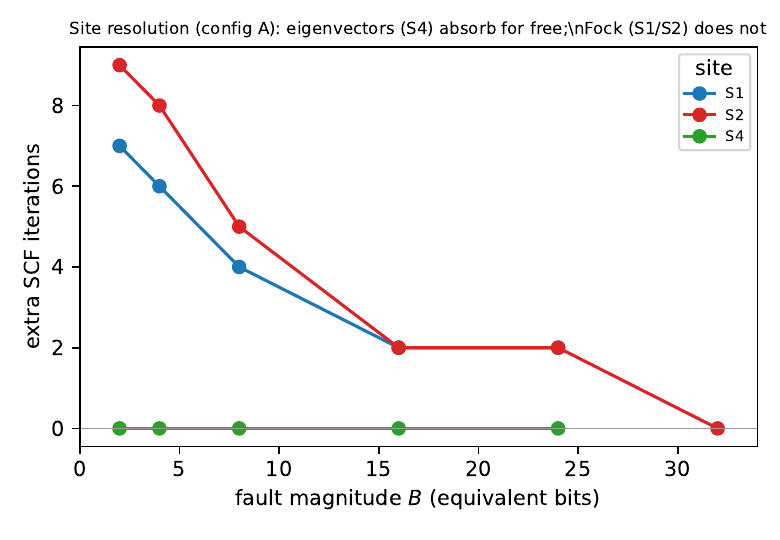}
\caption{Site resolution on CH$_2$ (production converger,
near-converged, adversarial element): Fock-side faults cost
iterations; orbital-matrix faults are nullified by occupied-space
projection.  On methanol the same orbital-site policy lands in the
occupied subspace and silently corrupts instead
(Sec.~\ref{sec:results}).}
\label{fig:sites}
\end{figure}
 
\begin{figure}[t]
\centering
\includegraphics[width=\columnwidth]{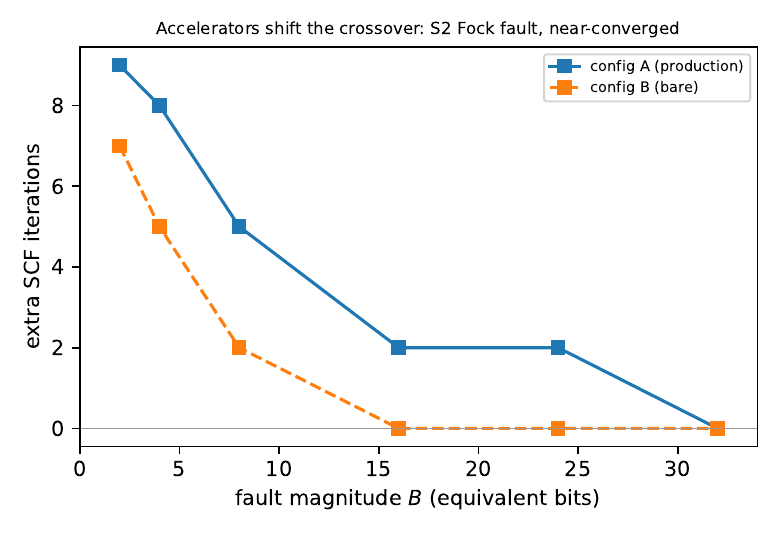}
\caption{Additional iterations under the production convergence
configuration and the bare fixed-point configuration for
near-converged Fock-side perturbations.  The production configuration
has a larger crossover value and an approximately constant
reconvergence cost over the tested range.}
\label{fig:avb}
\end{figure}
 
\subsection{Effect of SCF Acceleration on Fault Response}
Figure~\ref{fig:avb} compares fault absorption under the production
and bare fixed-point configurations. In the
production converger, the fitted slopes are shallow ($0.33$ at the
Fock site, $0.23$ at the density site) with crossovers near
$B^{*} \approx 26$--$29$. The fitted crossover for 
SOSCF is approximately sixteen bits larger
than that of the bare Roothaan iteration; above this crossover, 
the measured additional cost is approximately
constant. Two converse observations
complete the picture.  At gross magnitude the total delay of the production
configuration  \emph{exceeds} that of the bare one (9
vs.\ 7 iterations at $B=2$): past its absorption range, the
quasi-Newton machinery must rebuild its own perturbed model.  And
the early-phase arm inverts the ranking outright: iteration-3
faults add no iterations and, in several cells, reduce the iteration
count by one under bare iteration,
while the production converger requires an additional 3--5 iterations at
$B \le 8$.  The accelerator-internal state therefore introduces an additional
fault-sensitive component; the memoryless iteration has nothing to
corrupt.  Together, the experiments show that
acceleration state both absorbs faults and propagates them,
depending on what is corrupted.
 
\begin{figure}[t]
\centering
\includegraphics[width=\columnwidth]{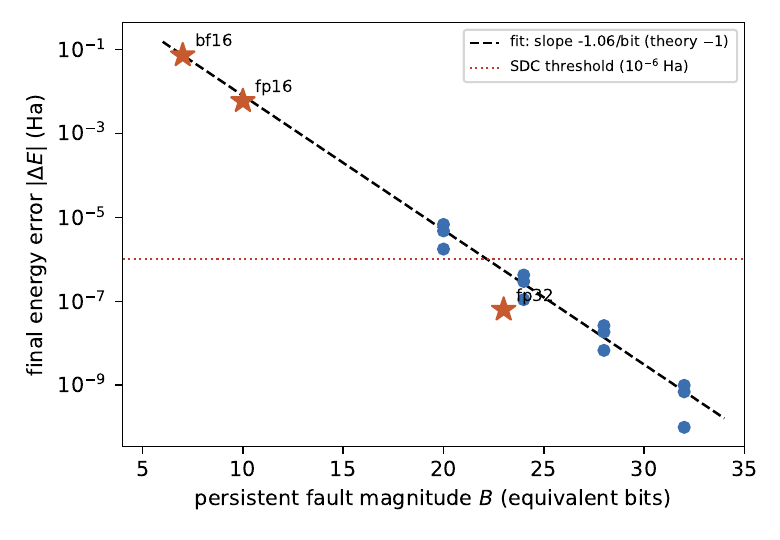}
\caption{Final-energy error under persistent Fock perturbations and
whole-array Fock quantization.  The persistent relative-noise results
have a fitted log-scale slope of $-1.06$ per bit and cross the
$10^{-6}$\,Ha threshold near $B=22.2$.}
\label{fig:sticky}
\end{figure}
 
\subsection{Persistent Faults and Reduced Precision}
Figure~\ref{fig:sticky} tests the persistent-fault corollary.
Sticky Fock noise produced a converged-energy floor with fitted
slope $-1.06$ per bit against the registered $-1.00$, approximately four orders of 
magnitude in $|\Delta E|$ over the
tested twelve-bit interval, crossing the SDC threshold at
$B^{*} = 22.2$ bits.  The storage-precision experiments exhibit a similar ordering: 
fp32 Fock storage every iteration converges
to the correct answer at a cost of two iterations
($|\Delta E| = 6\times 10^{-8}$\,Ha); fp16 silently misconverges
by $5.9$\,mHa; whereas the bf16-style truncation produces false convergence, 
declaring convergence \emph{seven iterations early} with
a 73\,mHa error (about 46\,kcal/mol) and normal termination.
Quantization to eight bits of significand precision reduces the
observed density-change signal until the two-pass convergence test is satisfied after 
the density-change values reach the quantization
scale: reduced precision does not slow this solver down; it
manufactures false convergence.
 
\subsection{Parallel Faults and Replica Coherence}
Single-rank Fock faults in 4-way runs (production configuration,
adversarial element, master rank) exhibit a different magnitude dependence from the serial runs.  Large faults ($B \le 8$) delayed uniformly, all
phases, fifteen of fifteen per magnitude, and never hung. 
These perturbations modify the globally reduced state sufficiently
that all replicas follow the same altered trajectory.  
Hangs appeared only at the
smallest magnitude in the sweep, $B=24$, and only in the
near-converged arms: 4 of 5 seeds in each phase-triggered cell,
zero at iteration 3.  A $2^{-24}$ fault is too small to move the
shared trajectory; it survives as a private offset in the state of the faulted
rank state, and at the production-converger superlinear
endgame, where the density change falls through the convergence
threshold in about one iteration, a private offset decides which
iteration a replica exits.  One rank leaves the loop; its peers
post another collective; the run deadlocks.  Within this campaign, 
the deadlock-producing perturbations are
therefore smaller than those that measurably alter the shared
trajectory.  Two follow-ups
sharpened the mechanism.  Forcing the convergence scalar of the master 
to zero under the production converger was absorbed both times: at
that converger cliff, every threshold crossing coincides with
genuine convergence, so the disagreement window has zero width.
This result did not support the corresponding predicted deadlock
mechanism.
The same fault under the bare converger, where linear contraction
spreads the crossings, instead produced an early false convergence
on the master that \emph{propagated}: a silent corruption born
from a convergence-scalar fault, demonstrating a distinct S5 failure pathway.
 
\begin{figure}[t]
\centering
\includegraphics[width=\columnwidth]{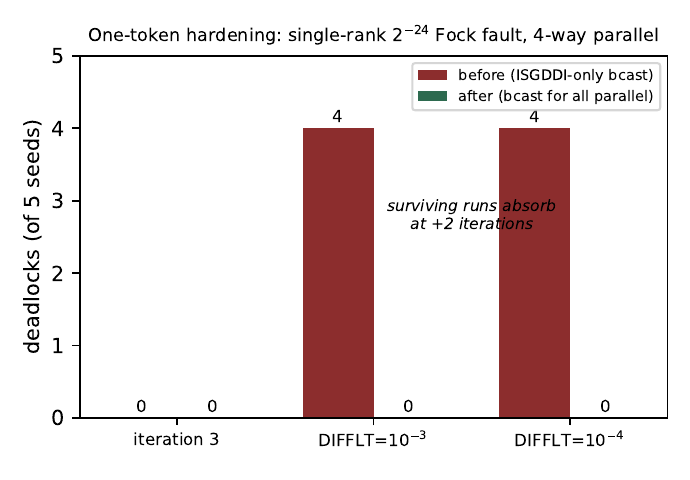}
\caption{Parallel fault outcomes before and after extending the
convergence-scalar broadcast from the master rank to all parallel runs.
The synchronization change eliminates the observed
replica-divergence deadlocks.  Runs that previously deadlocked
instead complete with two additional iterations, while the
downstream scalar-corruption campaigns reproduce their original
outcomes for every seed.}
\label{fig:hard}
\end{figure}
 
\subsection{Evaluation of a Convergence-Synchronization Change}
\label{sec:hardening}

Motivated by the replica-divergence results in
Sec.~\ref{sec:prelim}, we extended the existing broadcast of the
master-rank convergence scalars from the GDDI-specific execution
path to all parallel runs.  This modification changes the governing
condition to \texttt{GOPARR} but otherwise leaves the convergence
logic unchanged.  We compiled the modified implementation as a
separate binary and repeated the parallel divergence campaign shown
in Fig.~\ref{fig:hard}.

The synchronization change reduces the observed deadlock count from
8 of 10 affected runs to 0 of 10.  Each formerly deadlocking run
instead completes with two additional iterations, because all
replicas now follow the convergence decision communicated by the
master rank.  The unfaulted parallel baselines remain identical to
the corresponding serial results.

The scalar-corruption experiments reproduce their pre-change
outcomes for every seed, including the single-seed hang and the
false-convergence outcome under the bare fixed-point configuration.
These injection sites occur downstream of the broadcast and are
therefore not expected to be affected by the modification.  Within
the tested campaign, the change consequently eliminates the
identified replica-divergence mechanism without altering outcomes
outside its intended scope.  It also makes the master-rank
convergence scalars authoritative, concentrating the residual risk
of scalar corruption at that rank.
 
\begin{figure}[t]
\centering
\includegraphics[width=\columnwidth]{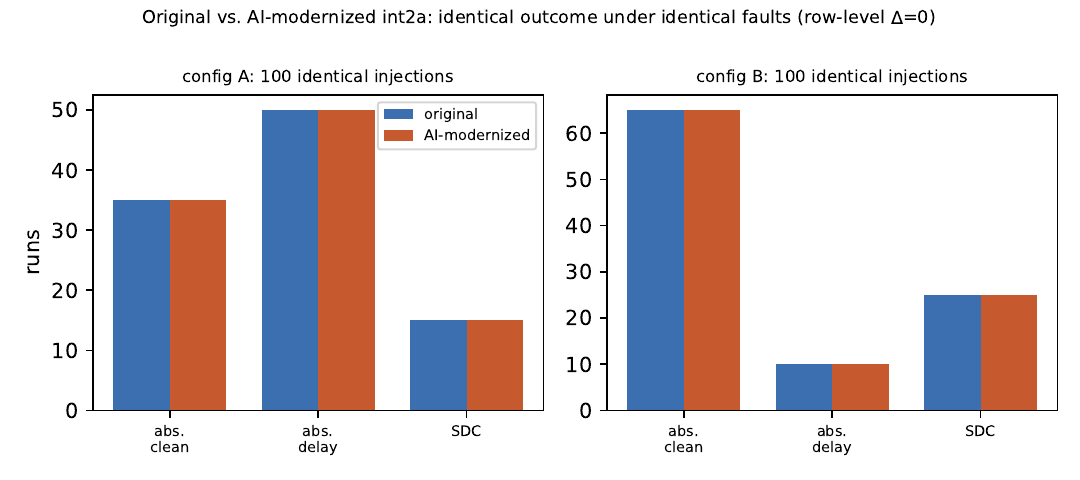}
\caption{Outcome counts for 100 paired injections per SCF
configuration using the original and AI-modernized
\texttt{int2a} implementations.  The paired runs agree in status,
iteration count, injection count, and final-energy representation.}
\label{fig:diffres}
\end{figure}
 
\subsection{Differential Comparison of Original and AI-Modernized Kernels}
Before the fault campaign, the modernized kernel was evaluated using
the following sequence:
canonical-flag compile; symbol parity with the legacy object; a
digit-for-digit reproduction of the standard geometry-optimization
battery on its first in-application execution (seven SCF cycles,
energies to all ten printed decimals, iteration counts identical),
notable because that the conventional path of the battery exercises the
one semantic edit of the strip; and clean-path nulls on the
direct-SCF, Rys-forced decks that the campaign uses, where the
converted kernels perform all integral work.  The campaign then
ran 100 identical injections per configuration on each binary,
density and Fock sites, five magnitudes, two phases, five seeds.
Figure~\ref{fig:diffres} summarizes the paired comparison.  No
disagreements were observed among the 200 paired runs. 
Status, iteration count, and firing count are
identical in all 200 pairs, and the maximum final-energy
discrepancy is $0.000$, with identical binary representations.  
The agreement is nontrivial because the campaign includes both
absorbed faults and silent corruptions: the direct-SCF regime
retains faults through differential Fock formation, and the
campaign produced 15 silent corruptions in the production
configuration and 25 under the bare converger; the modernized code
reproduced every one, at the same magnitudes, phases, and seeds,
with identical final-energy representations.  Within the tested fault 
space, the paired results provide no evidence
that the modernization changes the off-nominal behavior of the kernel.
This conclusion is limited to the selected sites, fault modes,
magnitudes, phases, input systems, and build configuration, but it
extends the clean-path validation to executions containing controlled
numerical perturbations.
 
\subsection{Combined Modernization and Hardening}
\label{sec:endstate}
We next applied the convergence-synchronization change to the
implementation containing the AI-modernized kernels and repeated the
40-cell divergence campaign.  The initial comparison produced
differences in 15 cells.  Examination of the per-run event logs showed
that the two source trees used different harness versions.  The event
formats differed, and the newer version used a modified seed-mixing
procedure, so equal nominal seeds did not produce identical element
selections.  The
per-run provenance logs identified a harness version skew between
the trees in a single table read (the two harness versions' event
formats differ, and the newer-version seed mixing changes draw
sequences at equal seeds).  The matched-harness rerun still
differed in 17 cells, now purely as one extra absorption iteration
in near-converged Fock cells: with draws provably identical (every
scalar-site arm reproduced exactly, the deterministic single-seed
hang included), the residual difference was the compiled object
itself, built under the validation flag set (\texttt{-O1}) rather
than the production set (\texttt{-O2} with its
floating-point-relevant options).  Rebuilt under production flags,
the composed binary matched the hardened original in \emph{40 of
40 cells, seed for seed}: statuses, iteration counts, firing
counts, and energy deviations identical, deterministic hang and
false-convergence corruption included, the entire battery
executing under floating-point exception trapping.  Two
conclusions follow.  The end state exists and is measured: an
AI-modernized \gamess{} identical to the original across all 
tested campaign cells  under identical faults and eliminates 
the observed failure channel without introducing differences 
in the tested comparison.  In this experiment, 
reduced precision does not merely increase the
iteration count; sufficiently coarse quantization causes the
convergence test to accept a substantially inaccurate state.
 
\section{Threats to Validity}
\label{sec:threats}
\emph{Fault-model realism.}  Application-level perturbation is a
proxy for hardware faults; we mitigate by spanning magnitudes from
single-bit-scale to gross corruption, including sticky modes
motivated by fleet studies~\cite{dixit2021silent}, and by showing
that the persistent-noise law and the storage-precision outcomes
fall on one curve.  Truncation modes chop rather than round;
relative noise leaves exact zeros unchanged; both deliberate, both
reported.  

\emph{Harness limitations and corrections.}
The campaigns surfaced two harness defects: correlated first draws
for adjacent seeds, which together with symmetry sparsity
degenerated the CH$_2$ random-element arm, and an effectiveness
ledger limited by 9-digit log formatting.  Both were fixed
(harness v1.1: seed mixing; 16-digit logging), both were confined
to bookkeeping arms, and the repaired random-element arm produced
live data on the C$_1$ molecule.  No reported absorption number
depends on either.  The composition run added two further instances,
both identified through comparison of the event logs and paired
outcomes: a harness version
skew between trees, identified in one table read from the
event-log formats, and a compilation-flag numerical skew,
identified as a systematic one-iteration shift at the convergence
cliff and eliminated by flag matching
(Sec.~\ref{sec:endstate}).  

\emph{Selection policies.}  The
largest-element policy is adversarial only where the largest
element is consequential; the S4 results shows the same policy
spanning harmless to catastrophic with molecular symmetry.  An
occupied-restricted variant is the registered refinement.

\emph{Scope.}  Two small closed-shell RHF systems, single node,
one application; the divergence and hardening arms used four
replicas.  The differential covers the principal converted file
under forced Rys routing; the sibling conversions are staged, with
their required bare-symbol surface already enumerated by the link
gate.  Rate extrapolation to production scale is out of scope; the
protocol, the law, and the mechanisms are the transferable claims.
 
\section{Conclusion}
\label{sec:conclusion}
This paper presents differential fault injection as a method for
evaluating LLM-modernized scientific software beyond nominal
regression tests.  The shared-driver harness introduces identical
faults into the original and modernized \gamess{} implementations,
allowing their responses to be compared without requiring
implementation-specific injection sites.  Across more than
2{,}200 runs, the campaigns characterize transient, persistent,
precision-related, orbital, and replica-local perturbations.  The
measured absorption costs for transient faults are consistent with a
contraction-based model evaluated at two contraction rates, while the
persistent-fault experiments exhibit an approximately factor-of-two
reduction in energy error per additional bit over the observed linear
regime.  The characterization also identifies false convergence under
coarse Fock quantization, convergence to alternate solutions following
large orbital perturbations, and a parallel deadlock caused by
rank-local disagreement near the convergence threshold.

The campaigns also demonstrate a use case of the validation procedure.
Specifically, the parallel campaign evaluates a
synchronization change suggested by an existing developer comment.
Within the tested campaign, this change eliminates the observed
deadlock channel without altering outcomes outside its intended scope.

For the primary differential comparison, the original and modernized
kernels agree in outcome classification, iteration count, injection
count, and final energy for all 200 paired injections.  After the
synchronization change is composed with the modernization and the
harness and compilation configurations are matched, the resulting
implementation also agrees with the hardened original in all 40
paired runs.  The experiments therefore distinguish two related but
separate objectives: demonstrating that a modernization preserves the
reference-implementation observed off-nominal behavior and
demonstrating that a targeted change improves a specific resilience
property.  Both objectives can be evaluated through reproducible,
paired fault-injection campaigns.
 
The validation procedure, rather than any application-specific
threshold or fault rate, is the principal transferable contribution.  
Next steps follow the enumeration the instrument already produced:
integrating the sibling conversions against the linker-derived
interface specification; extending sites to UHF, DFT, and
static-data corruption; occupied-restricted adversarial policies;
multi-node replica counts; and applying the same gates, unchanged,
to each further kernel the modernization delivers.  Because the harness 
hooks reside in code shared by both trees, each
future comparison can use the same paired design: identical injected
faults followed by a direct comparison of the resulting behavior.
 
\section*{Acknowledgments}
This research was supported in part by the U.S. Department of Education under the grant \# P116S230016 and in part by the Research Computing clusters at Old Dominion University.
 
\bibliographystyle{IEEEtran}
\bibliography{refs}
 
\end{document}